\documentclass[prb,aps,showpacs,unsortedaddress,superscriptaddress,amsmath,amssymb,floatfix,twocolumn]{revtex4-1}
\usepackage{graphicx}
\usepackage{epsfig}
\usepackage{graphicx}
\usepackage{amssymb}
\usepackage{amsmath}
\usepackage{wasysym}
\usepackage{hyperref}
\usepackage[usenames]{color}

\begin{document}

\title{X-Ray Photoemission Spectroscopy in the Falicov-Kimball model}
\author{Nandan Pakhira}
\affiliation{Department of Physics, Indian Institute of Technology, Kharagpur, West Bengal 721302, India.}
\affiliation{Department of Physics, Kazi Nazrul University, Asansol, West Bengal 713340, India.}
\author{A. M. Shvaika}
\affiliation{Institute for Condensed Matter Physics of the National Academy of Sciences of Ukraine, 1 Svientsitskii Street, 79011 Lviv, Ukraine}
\author{J. K. Freericks}
\affiliation{Department of Physics, Georgetown University, Washington, DC 20057, USA.}
\enlargethispage*{100pt}
\begin{abstract}
We calculate the finite temperature X-ray photoemission spectroscopy for the Falicov-Kimball model using a Weiner-Hopf sum equation approach. 
In the metallic state, the core-hole spectral function shows two side peaks corresponding to the creation of a core-hole on an empty site (or a doubly occupied site) and also has two nearly  degenerate central peaks (because of 
our choice of the model parameters) corresponding to the creation of a core-hole on a singly occupied site. The nearly doubly degenerate central peaks 
merge into a single peak at higher temperatures. In the insulating state, we obtain two peaks and a strongly temperature dependent low-energy peak corresponding 
to the creation of a core-hole on a thermally excited empty site. These results for the insulating state should be similar to those of the more 
general Hubbard model. Also, the strong correlations suggest that even without any additional broadening due to Auger like processes, the core-hole lifetime will be 
short.  
\end{abstract}


\maketitle
\section{Introduction}
In a large class of X-ray spectroscopic techniques like X-ray photoemission spectroscopy (XPS), X-ray absorption spectroscopy (XAS) and resonant inelastic X-ray 
scattering (RIXS) an incident high energy X-ray photon (typically with an energy $\sim 1-10\mbox{ keV}$) knocks an electron out of a deep core level 
state either out of the sample (XPS) or to an unoccupied state in the conduction band (XAS and RIXS) thereby creating a hole in the deep core level 
state. The conduction band electrons then feel a local attractive static potential due to the created core-hole and the system relaxes to a new 
orthogonal ground state through particle-hole excitations in the conduction band. As a direct consequence of this many body effect, the XAS and XPS show power-law divergences at the threshold energy when $T=0$. This phenomenon is known as the 
orthogonality catastrophe, as proposed by Anderson.~\cite{AndersonPRL1967} For noninteracting metals at zero temperature the 
exponent of the power law and the relative intensity~\cite{MahanPR1967,NozieresPR1969,DoniachJPC1970} of the XPS spectra are well understood. At finite 
temperature the power-law singularity is cut off by the thermal fluctuations. 

Much less is known about the fate of the power law divergence in the XPS spectra of strongly correlated metals and Mott insulators, although this question has been examined by others.\cite{LeePRL1992,MedenPRB1998,CornagliaPRB2007} The theory for strongly correlated XPS can be applied to XPS studies of transition-metal oxide 
compounds~\cite{HoribaPRL2004,KimPRL2004,TaguchiPRB2005,PanaccionePRL2006}, which are either Mott insulators or doped Mott insulators with interesting 
ground state magnetic properties. Cornaglia and Georges\cite{CornagliaPRB2007} have studied core-level photoemission spectra of the 
Hubbard model across the metal-insulator transition. The calculated XPS spectra in the metallic phase show an asymmetric power law divergence with an 
exponent that depends on both the Hubbard interaction $U$ and the core-hole potential $Q$. With increasing $U$, the exponent either vanishes continuously (when $Q$ is less than half the bandwidth) or remains nearly constant, but the weight under the peak vanishes (when $Q$ is more than half the bandwidth). This study was limited to zero temperature so it is not able to determine how the behavior changes as the temperature is raised (except to note that an actual power-law divergence only holds exactly at $T=0$).  These thermal effects become increasingly important for anticipated experiments with X-ray free-electron lasers, where pumping the system before probing it can lead the system to be in an effective high-temperature state.  This work would then be a first approximation to describing that complicated nonequilibrium phenomena by an effective high-temperature equilibrium theory.

We study the temperature dependence of core-hole spectral function in the Falicov-Kimball (FK) model,\cite{FalicovPRL1969} which is also the XPS spectral function because the core-hole level is so far below the Fermi level. The FK-model can be thought of as a 
special case of the more general Hubbard model in which one of the spin species (say the down spins) is static, while the other spin species (say the up spins) hop
through the annealed background of the static spin species with a nearest-neighbor hopping amplitude $t$. When two electrons of opposite spin are on the same lattice site, they interact with a Coulomb interaction  $U$.  Despite its simplicity, the FK-model has a metal-insulator transition for 
large Coulomb repulsion $U>U_{c}$ and is exactly solvable via dynamical mean-field theory (DMFT).\cite{freericks_zlatic_review} The local propagator for the itinerant species can be calculated exactly while the propagator for the static species can be 
calculated systematically by using numerical renormalization group~\cite{WilsonRMP1975,KrishnamurthyPRB,KrishnamurthyPRB2,BullaRMP2008,AndersPRL2005,AndersPRB2006,PetersPRB2006,WeichselbaumPRL2007} (NRG) or the Weiner-Hopf sum equation approach~\cite{McCoyBook,ShvaikaCMP2008,ShvaikaCMP2012} (at finite temperature). The most notable difference between the 
two models is that the metallic state of the FK-model is a non-Fermi liquid, while the metallic state in the Hubbard model is a Fermi liquid. The insulating 
state in both models has the same origin and the charge dynamics in the incompressible Mott insulating state is similar. This relationship between the two models in the insulating phase has been illustrated in studies of nonresonant Raman 
scattering.\cite{FreericksPRB2001,FreericksPRB2001s} 

We introduce an additional core-hole into the FK-model and study its finite temperature spectral properties using the Wiener-Hopf sum equation approach for various 
parameter strengths. This gives the XPS spectral function as well. 
The organization of the rest of the paper is as follows; In Sec. II, we introduce the core-hole problem, in Sec. III we introduce a mathematical formulation 
for calculation of the real time Green's function. In Sec. IV, we study the core-hole spectral function using the Wiener-Hopf sum equation approach and finally, in Sec. V, we 
conclude.       
\section{Core-hole problem in the Falicov-Kimball model}
The Falicov-Kimball~\cite{FalicovPRL1969} was originally proposed as a model for rare-earth compounds near a metal-insulator transition. It involves the interaction between mobile conduction 
$d$-electrons and the static localized $f$-electrons. The model can be applied to real rare earth compounds in an approximate way (in the \textit{incoherent} 
high temperature region).\cite{ybin4} The Hamiltonian for the 
Falicov-Kimball model (in the hole representation, with an additional core hole) is given by
\begin{eqnarray}
\mathcal{H} & = & -\frac{t^{*}}{2\sqrt{D}}\sum\limits_{\langle ij \rangle} d_{i}^{\dagger}d_{j} - \sum_{i}\mu n_{di}+\sum_{i}(E_{f}-\mu)n_{fi}\nonumber \\
  & & +\sum_{i}(E_{h}-\mu)n_{hi} + \sum_{i}Un_{di}n_{fi}+\sum_{i} Q_{d} n_{di}n_{hi}\nonumber \\ 
  & & +\sum_{i}Q_{f}n_{fi}n_{hi},
\label{eq:Hcore-hole}
\end{eqnarray}
where $t^{*}/2\sqrt{D}$ is the nearest neighbor hopping amplitude of the itinerant $d$-holes on a $D$-dimensional hypercubic lattice and $\mu$ is the common 
chemical potential (we take the limit $D\to\infty$ and use $t^*$ as our energy unit). The symbols $n_{di}=d_{i}^{\dagger}d_{i}$, $n_{fi}=f_{i}^{\dagger}f_{i}$ and $n_{hi}=h_{i}^{\dagger}h_{i}$ are the occupation number operators for 
the $d$-hole, $f$-hole and the core-hole at a given site $i$, respectively. $U$ is the on-site repulsive Coulomb interaction between the itinerant $d$-hole 
and static $f$-hole, whereas $Q_{d}$ and $Q_{f}$ are the repulsive Coulomb interactions between the core-hole and the $d$-hole and $f$-hole, respectively. 
$E_{f}\sim 1\mbox{ eV}$ and $E_{h}\sim 0.1 - 10\mbox{ keV}$ are the site energies of the $f$ state and the core-hole state, respectively. The case of half-filling ($n_{f} = n_{d} = 0.5$) corresponds to the choice of $\mu=U/2$ and $E_{f} = 0$; this is the particle-hole symmetric case (in the restricted 
subspace involving $d$ and $f$ electrons only). Also, under a partial hole-particle transformation $d \rightarrow d^{\dagger}$ and $f \rightarrow f^{\dagger}$, the interaction between the core-hole and the $d$ and $f$ states transforms as $Q_{d} \rightarrow -Q_{d}$ and 
$Q_{f}\rightarrow -Q_{f}$, respectively, \textit{i.e.} becomes attractive instead and the core-hole energy $E_{h}$ gets shifted to $E_{h}+Q_{d}+Q_{f}$.

\begin{figure}[!htb]
\includegraphics[scale=1.0]{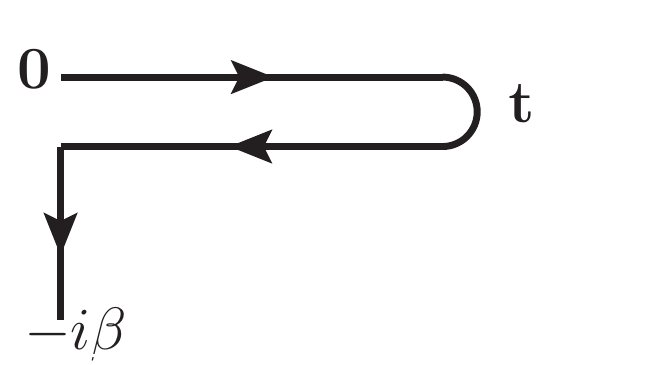}
\caption{The Kadanoff-Baym-Keldysh contour. The contour starts at time $t=0$, moves forward in time along the real axis to time $t$ then moves 
backward in time along the real axis to time $t=0$ and finally moves downwards along the imaginary axis to time $-i\beta$.}
\label{fig:KeldyshContour}
\end{figure}

Under DMFT, the model reduces to an effective single-impurity problem, described by the following local Hamiltonian  
\begin{eqnarray}
\mathcal{H}_{\textrm{loc}} = Un_{d}n_{f}+Q_{d}n_{d}n_{h}+Q_{f}n_{f}n_{h}-\mu n_{n}\nonumber \\
 +(E_{f}-\mu)n_{f}+(E_{h}-\mu)n_{h}
\label{eq:Hloc}
\end{eqnarray}
along with an effective time-dependent bath (arising from the degrees of freedom at all other sites except the site chosen) to which the $d$-holes hop in and out. 
The equilibrium density matrix for the single-impurity problem is given by
\begin{eqnarray}
\rho = \rho_{\textrm{loc}}\mathcal{T}_{c}\exp\left\{ - i\int_{c} dt' \int_{c} dt'' d^{\dagger}(t')\lambda_{c}(t',t'')d(t'')\right\},
\label{eq: density_matrix}
\end{eqnarray}  
where $\rho_{\textrm{loc}}=e^{-\beta H_{\textrm{loc}}}/\mathcal{Z}$, $\mathcal{Z}$ is the partition function (including the effects of the dynamical mean field), $\beta=1/k_{B} T$ is the inverse temperature 
and the time-ordering and integration are performed over the Kadanoff-Baym-Keldysh~\cite{Kadanoff-Baym,Keldysh} contour in Fig.~\ref{fig:KeldyshContour}. The time dependence of the operators in Eq.~(\ref{eq: density_matrix}) are given by the interaction representation with respect to $\mathcal{H}_{\textrm{loc}}$.

The time-translation-noninvariant dynamical mean field $\lambda_{c}(t,t')$ is given by 
\begin{equation}
\lambda_{c}(t,t')=-\frac{i}{\pi}\!\!\int\limits_{-\infty}^{+\infty}\!\!d\omega\textrm{Im}\left[\lambda(\omega)\right]e^{i\omega(t'-t)}\left[f(\omega)
-\Theta_{c}(t,t')\right],
\end{equation}
where $f(\omega)=1/[1+\exp(\beta\omega)]$ is the Fermi-Dirac distribution function and $\Theta_{c}(t,t')$ is the Heaviside function on the contour which is equal to 1 when $t$ 
is ahead of $t'$ on the contour, is equal to $0$ when $t$ is behind $t'$ and is equal to 1/2 when $t=t'$. Note that 
the dynamical mean field $\lambda(\omega)$ and the chemical potential $\mu$  are obtained from the equilibrium solution of the impurity problem without the
 core-hole. This in effect means that we are treating the creation of the core-hole under the sudden approximation instead of a fully self-consistent nonequilibrium 
treatment. The creation of the core-hole under the sudden approximation is commonly done because it is consistent with experiments.   

Because of the conserved core-hole number, $n_{h}$, and the conserved $f$-hole number, $n_{f}$, the full Hilbert space of the core-hole problem can be expressed 
as the direct sum of the Hilbert spaces in each conserved $\{n_{h},n_{f}\}$ sector. For the spinless case, the total partition function
for the single-impurity problem  $\mathcal{Z}$ contains four terms $\mathcal{Z}_{\alpha}\equiv\mathcal{Z}_{n_{h}n_{f}}$, each of which corresponds to the partition 
function for the Hilbert space in the conserved $\{n_{h},n_{f}\}$ sector. We have
\begin{eqnarray}
\mathcal{Z}=\mathcal{Z}_{00}+\mathcal{Z}_{01}+e^{-\beta(E_{h}-\mu)}\left[\mathcal{Z}_{10}+Z_{11}\right]
\end{eqnarray} 
with
\begin{eqnarray}
\mathcal{Z}_{00}&=&\left[1+e^{\beta\mu}\right]\prod_{m}\frac{i\omega_{m}+\mu-\lambda_{m}}{i\omega_{m}+\mu}\\
\mathcal{Z}_{01}&=&e^{\beta(\mu-E_{f})}\left[1+e^{\beta(\mu-U)}\right]\prod_{m}\frac{i\omega_{m}+\mu-U-\lambda_{m}}{i\omega_{m}+\mu-U}\\
\mathcal{Z}_{10}&=&\left[1+e^{\beta(\mu-Q_{d})}\right]\prod_{m}\frac{i\omega_{m}+\mu-Q_{d}-\lambda_{m}}{i\omega_{m}+\mu-Q_{d}}\\
\mathcal{Z}_{11}&=&e^{\beta(\mu-E_{f}-Q_{f})}\left[1+e^{\beta(\mu-U-Q_{d})}\right]\nonumber\\
&&\times\prod_{m}\frac{i\omega_{m}+\mu-U-Q_{d}-\lambda_{m}}{i\omega_{m}+\mu-U-Q_{d}},
\end{eqnarray}
where $i\omega_{m}=i\pi(2m+1)k_{B}T$ is the fermionic Matsubara frequency and 
\begin{eqnarray}
\lambda_{m}=\int_{0}^{\beta} d\tau \; e^{i\omega_{m}\tau}\lambda(\tau)
\end{eqnarray}
is the dynamical mean field evaluated at $i\omega_{m}$. The contour-ordered dynamical mean field $\lambda_{c}$ depends on the 
difference of its two time arguments when both of them lie on the imaginary time axis of the Kadanoff-Baym-Keldysh contour; we also use the notation 
$\lambda(\tau)=-i\lambda_{c}(-i\tau,0)$.
\section{Real time Green's functions}
We define the contour-ordered Green's function for the core-hole as
\begin{eqnarray}
G_{h}^{c}(t,t')=-i\langle\mathcal{T}_{c} \,h(t)h^{\dagger}(t')\rangle,
\end{eqnarray}
where the time ordering is taken along the Kadanoff-Baym-Keldysh contour shown in Fig.~\ref{fig:KeldyshContour} and $\langle\cdots\rangle$ corresponds to 
the trace weighted by the equilibrium density matrix in Eq.~(\ref{eq: density_matrix}). We also define the greater Green's 
function $G^{>}(t,t')=-i\langle h(t)h^{\dagger}(t')\rangle$ and lesser Green's function $G^{<}(t,t')=i\langle h^{\dagger}(t')h(t)\rangle$, which can all be expressed via
\begin{eqnarray}
G^{>}(t,t')=-\frac{i}{\mathcal{Z}}\sum_{m,n} e^{-\beta E_{m}}|\langle m|h|n\rangle|^{2} e^{i(E_{m}-E_{n})(t-t')}\\
G^{<}(t,t')=\frac{i}{\mathcal{Z}}\sum_{m,n} e^{-\beta E_{n}}|\langle n|h^{\dagger}|m\rangle|^{2} e^{i(E_{m}-E_{n})(t-t')}.
\end{eqnarray}
in the eigenbasis of the lattice Falicov-Kimball Hamiltonian with the additional core hole in Eq.~(\ref{eq:Hcore-hole}); the eigenstates satisfy $\mathcal{H}|n\rangle=E_n|n\rangle$. Note that we have suppressed the lattice site index in these equations, since the core-hole propagator is independent of the lattice site, but is always local, implying that the two hole creation and destruction operators must be from the same lattice site. Out of these two Green's functions we can construct retarded and 
advanced Green's functions,
\begin{eqnarray}
G_{h}^{r}(t,t') &=& -i\Theta(t-t')\langle\left[h(t),h^{\dagger}(t')\right]_{+}\rangle\\
                &=& \Theta(t-t')\left[G_{h}^{>}(t,t')-G_{h}^{<}(t,t')\right],\nonumber\\
G_{h}^{a}(t,t') &=& i\Theta(t'-t)\langle\left[h(t),h^{\dagger}(t')\right]_{+}\rangle\\
                &=& \Theta(t'-t)\left[G_{h}^{<}(t,t')-G_{h}^{>}(t,t')\right],\nonumber\\
\end{eqnarray}
respectively. The symbol $\left[ A,B\right]_{+}=AB+BA$ represents the anticommutator. 

Equilibrium problems are time-translation invariant, because there is no preferred time. This
is a property shared by the Green's functions, which follows by cyclic invariance of the trace and the fact that the lattice Hamiltonian commutes with itself. Hence, all of the Green's functions discussed here are functions only of $t-t'$. Furthermore, by relating the complex conjugate of a matrix element, to the Hermitian conjugate of the operators in the matrix element, we can also show that
\begin{eqnarray}
\left[G^{>}_{h}(t)\right]^{*}=-G^{>}_{h}(-t),\quad\left[G^{<}_{h}(t)\right]^{*}=-G^{<}_{h}(-t).
\label{eq:conjgG}
\end{eqnarray}
The core-hole spectral function $A_{h}(\omega)$ is then determined from
\begin{eqnarray}
A_{h}(\omega)=-\frac{1}{\pi}\textrm{Im}\left[G_{h}^{r}(\omega+i0^+)\right].
\label{eq:Ahdefn}
\end{eqnarray}

In many cases, one can directly perform an analytic continuation from Matsubara frequencies to real frequencies, but there is no obvious way to do that here.\cite{BrandtZPB1992,ZlaticPhilMagB2001,FreericksPRB2005}
Instead, one can always formulate the problem on the Kadanoff-Baym-Keldysh contour, and directly determine the Green's function as a function of time. It can then be Fourier transformed to frequency. This is the approach we adopt here. Furthermore,  for most core level states involving atomic $1s$ and $2p$ orbitals the core-hole energy is large $E_{h} \gg 0$ ($E_{h} \sim 600\textrm{ eV}$). In this limit, the presence of the thermal factor given by $\exp[-\beta(E_{h}-\mu)]$, completely suppresses the lesser Green's function, which we can approximate by zero. Hence, we have that the greater Green's function is identical to the retarded Green's function in this limit, namely $G_{h}^{r}(t)=\Theta(t)G_{h}^{>}(t)$. So we only need to compute
\begin{eqnarray}
G_{h}^{>}(t) &=& -i\textrm{Tr}\biggl[\mathcal{T}_{c}\exp\left\{-i\!\int_{c}\!dt'\!\int_{c}\!dt'' d^{\dagger}(t')\lambda_{c}(t',t'')d(t'')
\right\} \nonumber \\
&& \times h(t)h^{\dagger}(0)\rho_{\textrm{loc}}\biggr], 
\label{eq:Ghgreater}
\end{eqnarray} 
for $t\ge 0$.

We begin by solving the equations of motion for the core-hole operators, given by
\begin{eqnarray}
\frac{dh(t)}{dt} = -i\left[Q_{d}n_{d}(t)+Q_{f}n_{f}(t)+E_{h}-\mu\right]h(t),\\
\frac{dh^{\dagger}(\bar{t})}{d\bar{t}} = i\left[Q_{d}n_{d}(\bar{t})+Q_{f}n_{f}(\bar{t})+E_{h}-\mu\right]h^{\dagger}(\bar{t}),
\end{eqnarray}  
and substitute their solutions into Eq.~(\ref{eq:Ghgreater}), yielding
\begin{equation}
G^{>}_{h}(t)= -ie^{-i(E_{h}-\mu)t} \textrm{Tr}\left[e^{-\beta H_{0}}e^{-iQ_{f}n_{f}t} 
\;\mathcal{S}_{c}(t)h(0)h^{\dagger}(0)\right]
\end{equation}
where,
\begin{eqnarray}
\mathcal{S}_{c}(t)&=&\mathcal{T}_{c}\exp\left\{-i\int_{c}dt'\int_{c}dt'' d^{\dagger}(t')\lambda_c(t',t'')d(t'') \right. \nonumber \\ 
                  && \left. -i\int_{c} dt' Q_{c}(t,t')n_{d}(t')\right\},
\end{eqnarray}
with $Q_{c}(t,t')=Q_{d}$ for $t\in[0,t]$ \textit{on the upper branch of the contour only} and zero otherwise. It is due to the lack of time translation 
invariance of the $Q_{c}$ field that we must use the Kadanoff-Baym-Keldysh formalism for the analytic continuation. Note, however, that the final Green's function will remain time-translation invariant.

Since the core-hole occupation number $n_{h}=h^{\dagger}h$ is conserved, we have a projection onto states without a core-hole ($n_{h}=0$) only. Furthermore, since $n_f$ is also conserved, we add together the contributions for $n_f=0$ (top line) and $n_f=1$ (bottom line).  This results in 
\begin{eqnarray}
\label{eq:Ghgrt}
&&{G}_{h}^{>}(t) = -\frac{i}{\mathcal{Z}}e^{-i(E_{h}-\mu)t}\textrm{Tr}\left[e^{\beta \mu n_{d}}\mathcal{S}_{c}(t)\right] \\
                         &&  -\frac{i}{\mathcal{Z}}e^{-i(E_{h}-\mu)t}e^{-\beta(E_{f}-\mu)}e^{-i Q_{f}t}\textrm{Tr}\left[e^{\beta(\mu-U)n_{d}}\mathcal{S}_{c}(t)\right],\nonumber
\end{eqnarray}
where the trace is now over the $d$-holes only. 

The evaluation of the remaining traces is straightforward, because the actions are quadratic
in the $d$-electrons. But the steps one needs to follow are a bit involved. Following the methodology employed in the calculation of the $f$-particle propagator,\cite{ShvaikaCMP2008} we find
\begin{eqnarray}
\label{eq:Ggrtfinal}
{G}_{h}^{>}(t)&=&-ie^{-i(E_{h}-\mu)t}\left[\frac{\mathcal{Z}_{00}}{\mathcal{Z}}\mbox{det}_{[0,t]}
                (\mathbb{I}-Q_{d}G_{00})\right . \nonumber \\
     &+&\left .\frac{\mathcal{Z}_{01}}{\mathcal{Z}}e^{-iQ_{f}t}\mbox{det}_{[0,t]}(\mathbb{I}-Q_{d}G_{01})\right],
\end{eqnarray}  
where the Green's functions in the determinants are
\begin{eqnarray}
G_{n_h,n_f}(\omega)=\frac{1}{\omega+i\delta - \epsilon_{n_h,n_f}-\lambda(\omega+i\delta)},
\end{eqnarray}
in frequency space
with $\epsilon_{n_{h}n_{f}}$ given by $\epsilon_{00}=-\mu$, $\epsilon_{01}=U-\mu$,  $\epsilon_{10}=Q_{d}-\mu$, and $\epsilon_{11}=U+Q_{d}-\mu$. Note that the determinants 
in Eq.~(\ref{eq:Ggrtfinal}) are continuous matrix determinants over finite time intervals ranging from 0 to $t$. To determine the respective matrices in the time domain, we must recall that the identity matrix is given by a delta function and we must Fourier transform the real frequency Green's function back to the time representation.

The continuous time determinants are in the Toeplitz form because the Green's function $G_{n_h,n_f}(t_1,t_2)=G_{n_h,n_f}(t_1-t_2)$ depends only on the difference of the two time arguments. Asymptotic limits 
($t\rightarrow \infty$) of Toeplitz determinants can be easily calculated by using the Wiener-Hopf sum equation approach and Szeg\"{o}'s theorem~\cite{McCoyBook}. 
Two of us have also developed systematic finite-time corrections to these asymptotic forms.\cite{ShvaikaCMP2008} The finite time Toeplitz determinants 
can also be calculated through a discretization of the continuous matrix operators and a numerical integration over the Kadanoff-Baym-Keldysh contour. This latter method is found to be accurate in the 
small time region but becomes numerically intractable in the large time region because of the large matrices involved (due to the use of a fixed step size in time). We adopt 
a hybrid approach in which we use direct numerical evaluation of the Toeplitz determinants in the small time region and use Szeg\"{o}'s theorem and its finite time 
corrections to calculate the determinants in the large time region. This approach produces accurate real time Green's functions for all times.

\section{Core-Hole Spectral function}
We choose the familiar half-filling case which corresponds to $\langle n_{d}\rangle =\langle n_{f}\rangle=0.5$, $E_{f}=0$ and $\mu=U/2$, in the 
absence of the core-hole and as already mentioned we treat the creation of the core-hole under the sudden approximation; \textit{ i.e.} the creation of the core-hole 
does not modify the dynamics of the itinerant $d$-hole or static $f$-hole. We also choose hypercubic lattice, which in the limit of large dimensions ($D\rightarrow
\infty$) has a Gaussian density of states 
\begin{eqnarray}
D_{0}(\epsilon)=\frac{1}{\sqrt{\pi}}\exp{[-\epsilon^{2}]}
\end{eqnarray} 
where we set $t^*=1$. 
In the absence of the core-hole, the Falicov-Kimball model has a Mott insulating ground state for $U > U_{c}=\sqrt{2}$, when spatial ordering is suppressed. The core-hole spectral function $A_{h}(\omega)$ is also the XPS response function in the deep core-hole limit. We study the evolution of $A_{h}(\omega)$ for various temperatures $T$, interaction strengths $U$,  and 
core-hole potentials $Q_{d}$ and $Q_{f}$, keeping $Q_{d}=Q_{f}$ to reduce the number of parameters.  
\subsection{Weakly correlated metal}
\begin{figure}[!htb]
\begin{center}
\includegraphics[scale=0.32,clip=]{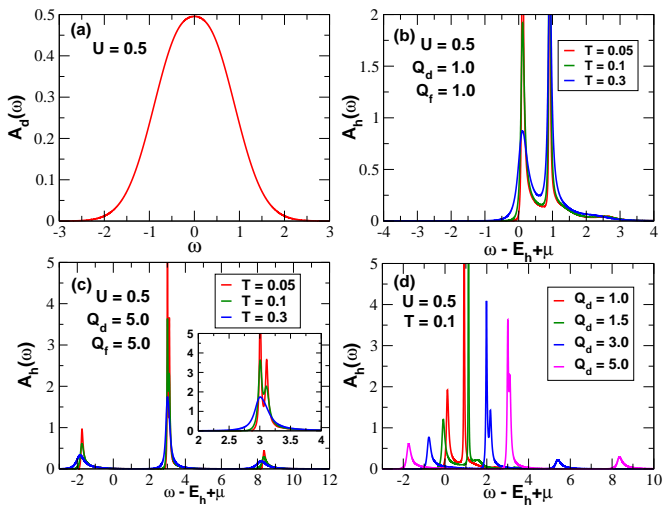}
\caption{(Color on-line). Panel (a): Itinerant species spectral function, $A_{d}(\omega)$. Panel (b) and (c): Temperature dependence of the core-hole spectral 
function, $A_{h}(\omega)$, for small ($Q_{d}=Q_{f}=1.0$) and large ($Q_{d}=Q_{f}=5.0$) core-hole potentials. Inset Panel (c): Blown up region near the central peak. 
Axis labels are the same as in the main panel. Panel (d): Evolution of $A_{h}(\omega)$ for various core-hole potentials $Q_{d}=Q_{f}$ for a given 
temperature $T=0.1$. Parameters used for the calculations are all indicated in the figures. }
\label{fig:SpectralFnU0.5}
\end{center}
\end{figure}

We first consider the case of a weakly correlated metal with $U=0.5$. As shown in Fig.~\ref{fig:SpectralFnU0.5} (a), the conduction electron spectral function $A_d(\omega)$ has a metallic
density of states, with no gap. In Fig.~\ref{fig:SpectralFnU0.5} (b) and (c), we show the temperature evolution of the core-hole spectral function for small 
$Q_{d}=Q_{f}=1.0$ and large $Q_{d}=Q_{f}=5.0$ core-hole potentials. The term small and large is with respect to $U$. 
$A_{h}(\omega)$ in Fig.~\ref{fig:SpectralFnU0.5}(b) shows two distinct peaks and a very broad hump whereas in Fig.~\ref{fig:SpectralFnU0.5}(c) we clearly see three 
distinct peaks. The peak structure can be easily understood from the atomic limit picture. Using the equation of motion, we can calculate the retarded core-hole 
Green's function as $\hat{G}_{h}^{r}(\omega)=\left(\omega+i\eta-E_{h}+\mu-Q_{d}\hat{n}_{d}-Q_{f}\hat{n}_{f}\right)^{-1}$, where one needs to insert the localized 
hole fillings for each sector. So, if we plot $A_{h}(\omega)$ as a function of $\omega-E_{h}+\mu$ then we will see two delta function peaks at $0$ and $Q_{d}$ for 
$n_{f}=0$ and two delta function peaks at $Q_{f}$ and $Q_{d}+Q_{f}$ for $n_{f}=1$. Because of the choice $Q_{d}=Q_{f}$ through out the calculation the two peak 
positions at $Q_{d}$ and $Q_{f}$ are degenerate. Note that the distance between two peaks in each of the conserved $n_{f}$ sectors is $Q_{d}$. Now once we couple the local Hamiltonian to the dynamical mean field, then each of the delta function peaks gets broadened due to finite lifetime of the atomic levels. Also, their peak 
position is shifted too. Most interestingly, the distance between the two peaks in each of the conserved $n_{f}$ sectors still remains equal to $\sim Q_{d}$. This 
is due to the fact that the core-hole--$d$-hole interaction shifts each of the $d$-hole levels by the same amount, namely $\sim Q_{d}$.  

In Fig.~\ref{fig:SpectralFnU0.5}(b), the peak near $\Omega\equiv\omega-E_{h}+\mu=0$ corresponds to the core-hole being created on a site not occupied by 
either a $d$- or $f$-hole. Whereas the peak at $\Omega \simeq 1.0$ corresponds to a nearly degenerate double peak, where the core-hole is
created on a site either occupied by a $d$ or $f$ hole. The broad shoulder at $\Omega \simeq 2.0$ corresponds to the case where the core-hole is created 
on a site doubly occupied by both a $d$ and $f$-hole. With increasing core-hole potentials $Q_{d}=Q_{f}$, the doubly degenerate central peak gets split as shown 
in the inset of Fig.~\ref{fig:SpectralFnU0.5}(c). This is due to the fact that local $d$-holes directly hybridize with the bath which shifts its energy 
and width, whereas the $f$-hole only indirectly sees the fluctuating bath through its interaction with the local $d$-hole;  therefore, it has different self-energy 
effects when compared to the $d$-hole. Thermal broadening again smears the two peaks and we obtain one broad central peak at high temperature ($T=0.3$).  

The integrated spectral weight under the central peak is much larger than that of the other two peaks. This corresponds to the fact that at half filling the excited 
core-hole would more probably be created at singly occupied sites rather than at unoccupied or doubly occupied sites. With increasing temperature the integrated 
weight under the two side peaks increases and that under the central peak decreases. This is because thermally excited sites create a pair of empty and doubly occupied sites at the expense of two singly occupied sites so the probability that the excited core-hole will be created at one of those sites increases with temperature.

In Fig.~\ref{fig:SpectralFnU0.5}(d), we show the systematic evolution of $A_{h}(\omega)$ at a fixed temperature $T=0.1$ for various core-hole potentials. The main 
 feature is that the peaks corresponding to the excitation of the core-hole onto empty or doubly occupied sites shifts farther and farther from the central peak.
It also appears that peaks are losing intensity, but since the integrated spectral weight is always equal to one, this is just an illusion---the change in peak heights is compensated by a change in their width.   

One may ask, where is the edge singularity? We can see that the edge singularity clearly disappears at nonzero temperature. As the temperature is lowered, one of the peaks becomes very sharp with a high amplitude and eventually it gives rise to a divergent signal with the power-law divergence occurring at $T=0$.\cite{ShvaikaCMP2012} Our focus here is on how this behavior changes at higher temperatures, and we can see that the behavior dramatically changes.
\subsection{Strongly correlated metal}
\begin{figure}[!htb]
\begin{center}
\includegraphics[scale=0.32,clip=]{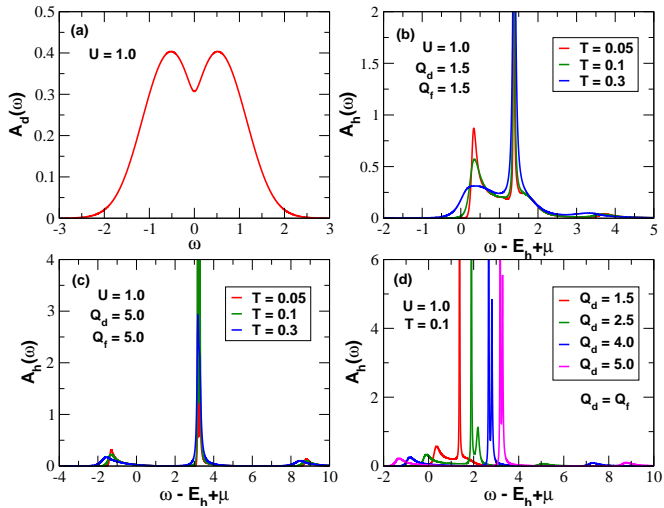}
\caption{(Color on-line) Panel (a) Conduction-electron spectral function with its pseudogap structure. Panel (b) and (c)  Temperature dependence of the core-hole 
spectral function for moderate ($Q_{d}=Q_{f}=1.5$) and large ($Q_{d}=Q_{f}=5.0$) core-hole potentials. Panel (d) : Evolution of the core-hole spectral function 
for various core-hole potentials $Q_{d}=Q_{f}$ at a fixed temperature $T=0.1$. The parameters used for the calculations are all indicated in the figures. }
\label{fig:SpectralFnU1.0}
\end{center}
\end{figure}

Next, we consider the case of a strongly correlated metal. We choose $U=1.0$. The conduction-electron density of states $A_{d}(\omega)$, is shown in 
Fig.~\ref{fig:SpectralFnU1.0}(a), and illustrates a pseudogap forming near the chemical potential.  
Figure~\ref{fig:SpectralFnU1.0} panel (b) and (c), plot the temperature evolution of the core-hole spectral function for moderately 
($Q_{d}=Q_{f}=1.5$) and large ($Q_{d}=Q_{f}=5.0$) core-hole potentials, respectively. One of the main features we see is that the central peak is narrowing and concentrating more spectral weight,
as the weight on the empty and doubly occupied sites goes down.
As we increase the core-hole potentials to large values the intensity of the central peak increases while its width decreases. This is due to increased 
self-energy effects arising when we put a core-hole onto a site already occupied by a $d$ or $f$-hole. The side peaks are well separated from the central peak and 
their intensity and integrated spectral weight under these peaks are reduced compared to the weakly correlated case. 
\subsection{Small-gap Mott insulator}
\begin{figure}[!htb]
\begin{center}
\includegraphics[scale=0.32,clip=]{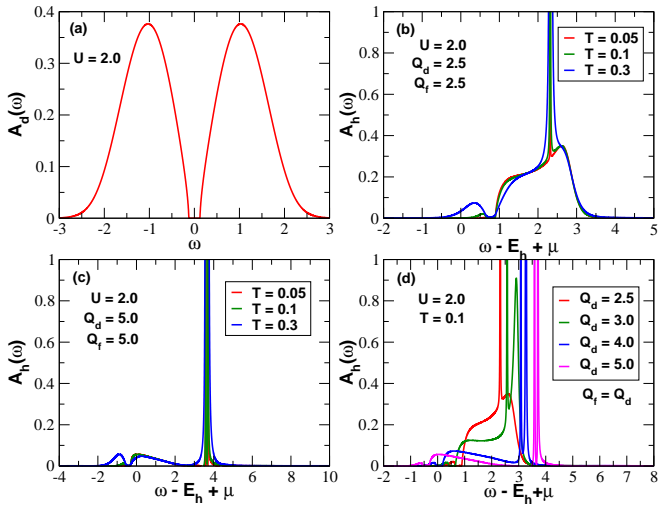}
\caption{(Color on-line) Panel (a) Conduction-electron spectral function, $A_{d}(\omega)$ in the  Mott insulator phase. Panels (b) and (c)  Core-hole 
spectral function, $A_{h}(\omega)$, for moderately large ($Q_{d}=Q_{f}=2.5$) and large ($Q_{d}=Q_{f}=5.0$) core-hole potentials. The peak corresponding to a
core-hole excitation onto the doubly occupied manifold has a small spectral weight. Panel (d) Evolution of $A_{h}(\omega)$ for various core-hole potentials 
$Q_{d}=Q_{f}$ at fixed temperature $T=0.1$. }
\label{fig:AhU2}
\end{center}
\end{figure}

As we increase $U$ above $U_{c}=\sqrt{2}$ we get into the Mott insulating phase. We choose $U=2.0$ which gives an insulator with a small gap, 
$\Delta_{\textrm{gap}} \simeq 0.25$. In panel (a) of 
Fig.~\ref{fig:AhU2}, we show the spectral function of the $d$-hole which clearly shows two Hubbard bands centered around $\pm\frac{U}{2}$ and separated by the 
insulating gap, $\Delta_{\textrm{gap}}$. Of course, it is well known that for the $D=\infty$ hypercubic lattice the Mott insulating density of states never has a true ``gap'' as there are always exponentially small density of states inside the ``gap region.''

Figure~\ref{fig:AhU2} panels, (b) and (c),  show the core-hole spectral function $A_{h}(\omega)$ for moderately large $Q_{d}=Q_{f}=2.5$ and large 
$Q_{d}=Q_{f}=5.0$ core-hole potentials. In the insulating state at zero temperature, a core-hole can only be created on a site which is occupied by either a 
$d$-hole or an $f$-hole. This gives rise to single peak in each of the two conserved sectors of $n_{f}=0$ and $n_{f}=1$. For $Q_{d}=Q_{f}=2.5$ we see a very 
sharp near $\delta$-function peak at $\Omega \sim Q_{f}$ on top of very broad asymmetrical spectral feature. The sharp peak corresponds to the creation of a core-hole 
on a site already occupied by an $f$-hole. The sharpness of the peak arises due to both the static nature of the $f$-hole and the lack of core-hole screening 
effects in the insulating state thereby making the core-hole state very long lived. The broad asymmetrical part of $A_{h}(\omega)$ corresponds to the creation 
of a core-hole on a site occupied by a $d$-hole. This part consists of a broad peak near $\Omega\sim Q_{d}$ on top of a broad background. As already 
mentioned, due to lack of screening effects in the insulating states and strong self-energy effects of the $d$-hole, the core-hole state is short 
lived. The width of the broad spectral feature is nearly equal to the width of 
the lower Hubbard band. At finite temperature, some of the sites become unoccupied while others becomes doubly occupied and creation of core-hole on 
either of these two sites gives rise to additional side band peaks. The peak corresponding to the  creation of a core-hole on an empty site corresponds the peak at 
$\Omega \sim -0.75$ separated by a temperature dependent pseudo-gap of the size $\Delta_{\textrm{gap}}$. The peak corresponding to the  creation of a core-hole on the 
doubly occupied site is extremely weak (nearly invisible on the scale of the plot) and has negligible spectral weight under it.   
 
With increasing core-hole potentials the peak corresponding to core-hole excitation into the $n_{f}=1$ manifold largely remains the same but the spectral features 
corresponding to a core-hole excitation onto the $n_{f}=0$ manifold shows qualitative changes as plotted in panels (c) and (d). With increasing $Q_{d}$, the broad peak 
at $\Omega \sim Q_{d}$ gets gradually pushed out to higher energy and gradually separates out of the broad background. Also, the core-hole lifetime 
increases, as is evident from the  narrowing and diverging peak at $\Omega \sim Q_{d}$. This is mainly due to the fact that the excited core-hole goes into the 
upper Hubbard band and its decay rate decreases with increasing $Q_{d}$ as the $d$-hole density of states involved in the decay process decreases with 
increasing $Q_{d}$. Due to the same reason, the integrated spectral weight under the broad background also decreases, while its width remains nearly same.  
\subsection{Large gap Mott insulator} 
\begin{figure}[!htb]
\begin{center}
\includegraphics[scale=0.32,clip=]{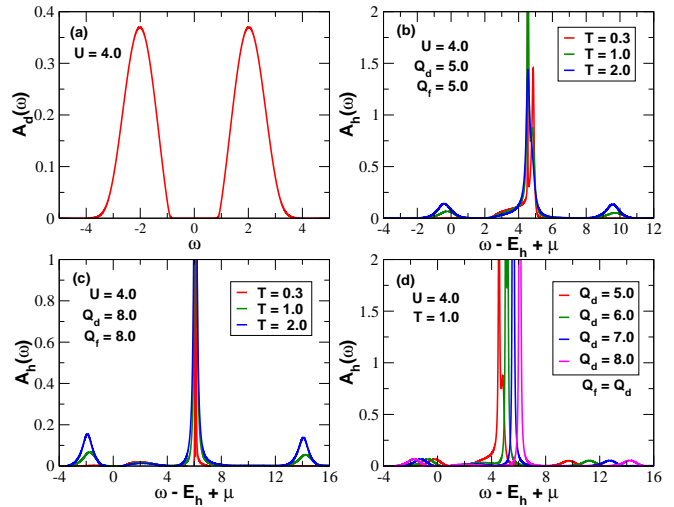}
\caption{(Color on-line) Panel (a) The $d$-hole spectral function $A_{d}(\omega)$ in a strong Mott insulator. Panel (b) and (c) Core-hole spectral function, 
$A_{h}(\omega)$, for large ($Q_{d}=Q_{f}=5.0$) and very large ($Q_{d}=Q_{f}=8.0$) core-hole potentials, respectively. For temperatures larger than 1, side peaks corresponding to core-hole excitations into empty and doubly occupied manifold are clearly visible. Panel (d)  Evolution of $A_{h}(\omega)$ 
with core-hole potentials $Q_{d}=Q_{f}$ for fixed temperature $T=1.0$. }
\label{fig:AhU4}
\end{center} 
\end{figure}

Finally we choose $U=4.0$ to simulate a strong Mott insulator. In panel (a) of Fig.~\ref{fig:AhU4}, we show the spectral function for the $d$-hole which clearly shows two Hubbard bands separated 
by a large insulating gap, $\Delta_{\textrm{gap}} \simeq 1.8$. In this case, the size of the insulating gap is much larger than the hopping. 
In Fig.~\ref{fig:AhU4} panels (b) and (c), we show the core-hole spectral function $A_{h}(\omega)$ for two core-hole potentials 
$Q_{d}=Q_{f}=5.0$ and $Q_{d}=Q_{f}=8.0$, respectively. Because of the presence of a large insulating gap in $d$-hole spectral function and strong correlation effects, the core-hole spectral functions shows negligible temperature dependence for $T \ll 1$. Therefore, we choose temperatures equal to or larger than 1. At the low temperature of $T = 0.3$, we see the two familiar spectral features corresponding to the core-hole being excited to two singly occupied manifolds. The peak at $\Omega \sim Q_{d}$ is much sharper and the broad spectral features have much less intensity compared to the case of small gap Mott insulator. This is 
mainly due to reduced density of states and increased correlation effects. For larger $Q_{d}=Q_{f}=8.0$, the broad spectral features have vanishingly small 
spectral intensity at $T=0.3$ as shown in panel (c). 

As we increase temperature equal to or larger than the hopping, the two nearly degenerate sharp peaks broaden and eventually merge into 
a single broad peak. Also, there is clear development of two additional spectral peaks at $\Omega \sim 0$ and $\Omega \sim Q_{d}+Q_{f}$, which  corresponds again to the creation of the core-hole on a thermally excited empty or doubly occupied site. 

In panel (d), we show the evolution of $A_{h}(\omega)$ with various core-hole potentials $Q_{d}=Q_{f}$ for $T=1.0$. With increasing core-hole potentials the central 
peaks becomes narrower and gradually merge into a well defined Lorentzian peak which separates out of the broad background. The side peaks move 
outward, but their distance from the central peak remains the same ($\sim Q_{d}$), as expected.  
\section{Conclusions}
In conclusion, we have calculated the finite temperature core-hole propagator (XPS spectrum) in the Falicov-Kimball model using the Weiner-Hopf sum equation approach. We have studied the
core-hole spectral function, $A_{h}(\omega)$ for various interaction strengths $U$ and core-hole 
potentials $Q_{d}=Q_{f}$. 

While the features of the weakly correlated metal are what we would expect at nonzero temperature---they have the power-law singularity suppressed and the spectral features broadened, as we enter the Mott phase, the system continuously evolves into a different type of spectral function. This is because the Mott insulator strongly suppresses doubly occupied and empty sites, hence the peaks associated with hole creation on those sites is sharply reduced. We also do not see any clear indication that there is a singularity at $T=0$ anymore (but we did not study this point in detail).

If we use these results to try to predict what the behavior of time-resolved XPS spectra would look like, we need to be aware of a few issues. First, we would expect the generic broadening of features and enhancement of the satellite peaks as the system absorbs more energy from the light pulse. Second, we also anticipate significant broadening of the sharp peaks in the spectra due to the probe pulse widths. Nevertheless, there are a number of interesting results one can predict could be extracted from such data. This includes the interaction energies $U$, $Q_d$, and $Q_f$, which can be extracted by examining the separations of different peaks. Since the spectral line shapes change so much with temperature, they might be able to be used as effective thermometers for the hot electron gases. Finally, an analysis of the weights in the satellite bands will yield information about the densities of empty and doubly occupied sites, which may be one of the more direct ways to measure doublon occupancy in the system.

What implications does this work have for more general models like the Hubbard model? The metallic phases are difficult to compare when we are below the coherence temperature of the Hubbard model, because the Falicov-Kimball model is not a Fermi liquid. But at higher temperatures, or in the insulating phase, we expect the behavior to be more similar.   We already know from Raman scattering studies that the charge dynamics in the insulators are similar.\cite{FreericksPRB2001,FreericksPRB2001s} The nearly delta function peak corresponding to the creation of a core-hole on a site occupied 
by a static $f$-hole will certainly be absent in the case of Hubbard model (instead it will be broadened like the $d$-hole contribution). So, we do expect that the temperature-dependent broad asymmetrical peak together with the small low 
energy peak corresponding to creation of a core-hole on a thermally excited empty site will survive in the Hubbard model. Also, the broad nature 
of this peak suggests that even without any additional broadening due to Auger-like nonradiative processes, the core-hole lifetime will be short. 
\section*{Acknowledgements}
 The work at Georgetown was supported by the Department of Energy, Office of
Basic Energy Sciences, Division of Materials Sciences
and Engineering (DMSE) under contract No. DE-FG02-08ER46542 (JKF and NP). 
NP also acknowledges financial support from IIT, Kharagpur where part of the manuscript was written.
JKF also acknowledges support from the McDevitt Bequest at
Georgetown. Fig.~\ref{fig:KeldyshContour} in this 
paper is created using open source software JAXODRAW and the original reference has been duly cited in Ref.~\onlinecite{Jaxodraw}.


\end{document}